\documentclass[twocolumn]{jpsj3}

\def\nn{\nonumber}

\newcommand{\bsigma}{\mbox{\boldmath $\sigma$}}

\usepackage{amsmath}
\usepackage{graphicx}

\title{Identifying the Orientation of Edge of Graphene Using G band
Raman Spectra}

\author{
Ken-ichi \textsc{Sasaki}$^{1}$
\thanks{E-mail address: SASAKI.Kenichi@nims.go.jp},
Riichiro \textsc{Saito}$^{2}$,
Katsunori \textsc{Wakabayashi}$^{1,3}$,
and
Toshiaki \textsc{Enoki}$^{4}$
}

\inst{
$^{1}$ International Center for Materials Nanoarchitectonics, \\
National Institute for Materials Science, Namiki, Tsukuba 305-0044,
Japan \\
$^{2}$ Department of Physics, Tohoku University, Sendai 980-8578, Japan \\
$^{3}$ PRESTO, Japan Science and Technology Agency,
Kawaguchi 332-0012, Japan \\
$^{4}$ Department of Chemistry, Tokyo Institute of Technology,
Ookayama, Meguro-ku, Tokyo 152-8551, Japan
}

\recdate{\today}

\abst{
 The electron-phonon matrix elements relevant to 
 the Raman intensity and Kohn anomaly of the G band are calculated 
 by taking into account the effect of the edge of graphene.
 The analysis of the pseudospin reveals that 
 the longitudinal optical phonon mode undergoes a strong Kohn anomaly
 for both the armchair and zigzag edges, and that only 
 the longitudinal (transverse) optical phonon mode 
 is a Raman active mode near the armchair (zigzag) edge. 
 The Raman intensity is enhanced 
 when the polarization of the incident laser light 
 is parallel (perpendicular) to the armchair (zigzag) edge. 
 This asymmetry between the armchair and zigzag edges 
 is useful in identifying the orientation of the edge of graphene.
}

\kword{graphene, edge orientation, Raman spectroscopy, electron-phonon
interaction, electron-light interaction, pseudospin, gauge field}

\begin{document}

\maketitle

\section{Introduction}

Graphene is a unique material
since its electron motion is governed by a special equation 
similar to the relativistic massless Dirac
equation, while a nonrelativistic equation is common
in condensed matter physics.~\cite{novoselov05,zhang05} 
The electron motion is modified by 
the electron-phonon (el-ph) and electron-light interactions,
which are fundamental issues in discussing 
the transport,~\cite{novoselov05,zhang05}
electronic,~\cite{bostwick07}
and optical properties~\cite{ferrari06,yan07}
of graphene.
The goal of this paper is to show that
an asymmetry of the Raman spectra for 
$\Gamma$ point longitudinal and transverse 
optical phonon (LO and TO) modes, 
both of which are known as the Raman G band,
appears near the edge of graphene.
There are two fundamental orientations for the edge of graphene,
zigzag and armchair edges, and a general edge shape
is considered to be a mixture of them.~\cite{kosynkin09,jiao09} 
The asymmetry is useful 
in identifying the orientation of the edge of graphene
by Raman spectroscopy.

In Raman spectroscopy, 
we irradiate laser light onto a sample
and observe the intensity of the inelastically scattered light.
The energy difference 
between the incident laser and the inelastically scattered light
corresponds to the energy of a Raman active phonon mode
due to the energy conservation.
The el-ph interaction is essential for the Raman process.
Further, the el-ph interaction can modify
the energy and life-time of the phonon mode,
which is known as the Kohn anomaly.~\cite{kohn59}
Evidence for Kohn anomalies is found in the phonon dispersion
of carbon nanotube,~\cite{dubay02}
graphene,~\cite{ando06-ka,lazzeri06prl}
and graphite.~\cite{piscanec04}
By examining the Kohn anomaly for the G band of carbon
nanotube,~\cite{farhat07} 
a feature of the el-ph interaction such as 
the chirality dependence of the el-ph interaction upon 
the Kohn anomaly has been clarified.~\cite{sasaki08_curvat}
In this paper, we calculate the el-ph matrix elements 
relevant to the Raman intensity and Kohn anomaly of the G band
of graphene within effective-mass approximation
by including the effects of the edge of graphene
and polarization direction of an incident laser (and a scattered) light.

This paper is organized as follows.
In \S~\ref{sec:hamiltonian}, 
we show the Hamiltonian including the el-ph interaction with respect
to the $\Gamma$ point optical phonon modes and the electron-light
interaction.
In \S~\ref{sec:zig} and \S~\ref{sec:arm}, 
we calculate the matrix elements for the el-ph and electron-light
interactions by taking into account of the presence of the zigzag and
armchair edges, respectively. 
The self-energy of the LO mode is estimated in \S~\ref{sec:selfene}
and the phonon self-energy for general edge shape is discussed. 
Finally, we propose two models representing the electronic states at the
interior of a graphene sample and calculate the self-energies for those
models in \S~\ref{sec:bulk}. 
In \S~\ref{sec:discussion}, we discuss the relationship between our
result and experimental results, and summarize the results.

\section{Hamiltonian}\label{sec:hamiltonian}

Let $\Psi_{\rm K}({\bf r})$ [$\Psi_{\rm K^\prime}({\bf r})$]
be the wave function for an electron near the K [K$'$] point,
the energy eigen equation for an electron near the Fermi
energy of graphene is written as 
\begin{align}
 {\hat H}
 \begin{pmatrix}
  \Psi_{\rm K}({\bf r}) \cr \Psi_{\rm K^\prime}({\bf r})
 \end{pmatrix}
 = E
 \begin{pmatrix}
  \Psi_{\rm K}({\bf r}) \cr \Psi_{\rm K^\prime}({\bf r})
 \end{pmatrix}.
\end{align}
The wave function $\Psi_{\rm K}({\bf r})$ 
[$\Psi_{\rm K^\prime}({\bf r})$]
is two-component structure, which
results from that the hexagonal unit cell contains two carbon atoms
[A atom ($\bullet$) and B atom ($\circ$) in Fig.~\ref{fig:zigLOTO}].
The total Hamiltonian ${\hat H}$ 
including the el-ph interaction 
with respect to the $\Gamma$ point LO and TO modes,
and the electron-light interaction 
is given by~\cite{sasaki08ptps}
\begin{align}
 {\hat H} = v_{\rm F}
 \begin{pmatrix}
  \bsigma \cdot ({\bf {\hat p}}+{\bf A}^{\rm q}-e{\bf A}) & 0 \cr
  0 & \bsigma' \cdot ({\bf {\hat p}}-{\bf A}^{\rm q}-e{\bf A})
 \end{pmatrix}.
 \label{eq:HKK'}
\end{align}
Here $v_{\rm F}$ is the Fermi velocity,
momentum operator ${\bf {\hat p}}=-i\hbar(\partial_x,\partial_y)$,
$\bsigma \equiv (\sigma_x,\sigma_y)$ and
$\bsigma' \equiv (-\sigma_x,\sigma_y)$ where
$\sigma_x$, $\sigma_y$ and $\sigma_z$ are 
Pauli matrices.
We take $x$ and $y$ axes as shown by the inset 
in Fig.~\ref{fig:zigLOTO}(a).
The electromagnetic gauge field ${\bf A}$
enters into the Hamiltonian through the substitution
${\bf {\hat p}} \to {\bf {\hat p}}-e{\bf A}$
where $-e$ is the charge of electron.
A uniform field ${\bf A}$ can represent 
the incident laser light and the scattered light
in the Raman process.
The el-ph interaction is represented by 
the deformation-induced gauge field
${\bf A}^{\rm q}=(A^{\rm q}_{x},A^{\rm q}_{y})$.~\cite{sasaki08ptps}
It can be shown that
$A^{\rm q}_{x}$ and $A^{\rm q}_{y}$ are expressed 
in terms of a change of the nearest-neighbor hopping integral
from an average value $-\gamma_0$, $\delta \gamma_{0,a}$,
as~\cite{kane97,sasaki05,katsnelson08}
\begin{align}
 \begin{split}
  & v_{\rm F} A^{\rm q}_x = \delta \gamma_{0,1}
  - \frac{1}{2} \left(
  \delta \gamma_{0,2}+ \delta \gamma_{0,3} \right), \\
  & v_{\rm F} A^{\rm q}_y = \frac{\sqrt{3}}{2} 
  \left( \delta \gamma_{0,2}- \delta \gamma_{0,3} \right).
 \end{split}
 \label{eq:gauge}
\end{align}
Here $a$ $(=1,2,3)$ for $\delta \gamma_{0,a}$ 
denotes the direction of the bond
(see the inset of Fig.~\ref{fig:zigLOTO}), and 
$\delta \gamma_{0,a}$ is caused by 
atomic displacements 
by the $\Gamma$ point optical phonon modes.
Note that ${\bf A}^{\rm q}$ is uniform 
for the $\Gamma$ point ${\bf q}=0$ phonons, 
while ${\bf A}^{\rm q}$ depends on the position ${\bf r}$
as ${\bf A}^{\rm q}({\bf r})$ for phonons with ${\bf q} \ne 0$.~\cite{sasaki08_curvat}
Although an additional deformation-induced gauge field 
due to a local modulation of the hopping integral 
originating from a defect appears in a realistic situation, 
we ignore it in eq.~(\ref{eq:HKK'}) for simplicity.

\section{Zigzag Edge}\label{sec:zig}

\begin{figure}[htbp]
 \begin{center}
  \includegraphics[scale=0.4]{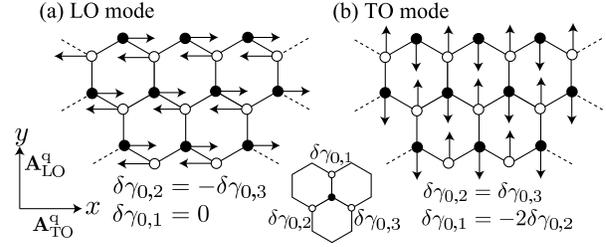}
 \end{center}
 \caption{The displacement vectors for the LO and TO modes 
 are shown in (a) and (b), respectively. 
 The displacement vectors of the LO (TO) mode 
 are parallel (perpendicular) to the zigzag edge.
 The direction of the deformation-induced gauge field ${\bf A}^{\rm q}$
 is perpendicular to the direction of atom displacement.
 }
 \label{fig:zigLOTO}
\end{figure}

First, we calculate the matrix element 
relevant to the Raman intensity near the zigzag edge.
The scattering or reflection of an electron at the zigzag edge
is intravalley scattering,~\cite{pimenta07} and therefore
we can consider the K and K$'$ points separately.
Let us consider the electrons near the K point.
The Hamiltonian is given by
\begin{align}
 {\hat H}_{\rm K}=v_{\rm F}
 \bsigma \cdot \left(
 {\bf {\hat p}}+{\bf A}^{\rm q}-e{\bf A} \right).
 \label{eq:HK}
\end{align}
We specify the deformation-induced gauge field 
${\bf A}^{\rm q}$ for the LO and TO modes near the zigzag edge.
The vibrations of carbon atoms 
corresponding to the $\Gamma$ point LO and TO modes 
are shown in Figs.~\ref{fig:zigLOTO}(a) and~\ref{fig:zigLOTO}(b).
By assuming that the perturbation
$\delta \gamma_{0,a}$ is proportional to 
the change in the bond length,
we have $\delta \gamma_{0,1} = 0$ and 
$\delta \gamma_{0,2}=-\delta \gamma_{0,3}$
for the LO mode, while
$\delta \gamma_{0,1} = -2 \delta \gamma_{0,2}$ 
and $\delta \gamma_{0,2}=\delta \gamma_{0,3}$
for the TO mode.
Using eq.~(\ref{eq:gauge}), we see that
${\bf A}^{\rm q}$ for the LO mode
is written as ${\bf A}^{\rm q}_{\rm LO}=(0,A_y^{\rm q})$
with $v_{\rm F}A_y^{\rm q}=\sqrt{3}\delta\gamma_{0,2}$, while 
${\bf A}^{\rm q}$ for the TO mode
is written as ${\bf A}^{\rm q}_{\rm TO}=(A_x^{\rm q},0)$
with $v_{\rm F}A_x^{\rm q}=-3\delta\gamma_{0,2}$.
Note that the direction of ${\bf A}^{\rm q}$ for the LO (TO) mode 
is perpendicular (parallel) to the zigzag edge.
The direction of ${\bf A}^{\rm q}$ 
is perpendicular to the direction of atom
displacement.~\cite{dubay02,ishikawa06}
Thus, the el-ph interaction in eq.~(\ref{eq:HK}),
$H^{\rm zig}_{\rm LO/TO}\equiv v_{\rm F} \bsigma 
\cdot {\bf A}^{\rm q}_{\rm LO/TO}$,
is rewritten as
\begin{align}
\begin{split}
 & H^{\rm zig}_{\rm LO} = v_{\rm F} A^{\rm q}_y \sigma_y, \\
 & H^{\rm zig}_{\rm TO} = v_{\rm F} A^{\rm q}_x \sigma_x,
\end{split}
\label{eq:Hzig}
\end{align}
for the LO and TO modes, respectively.

\begin{figure}[htbp]
 \begin{center}
  \includegraphics[scale=0.4]{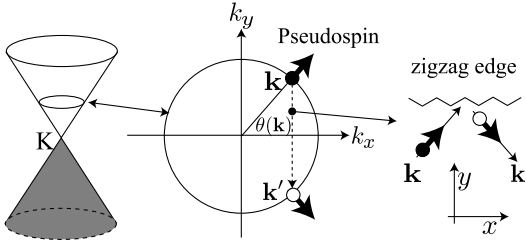}
 \end{center}
 \caption{
 The zigzag edge reflects the wave vector 
 ${\bf k}=(k_x,k_y)$ to ${\bf k'}=(k_x,-k_y)$, and
 two wave functions $\Psi^{\rm c}_{\bf k}({\bf r})$ 
 and $\Psi^{\rm c}_{\bf k'}({\bf r})$ form a standing wave.
 Note that the $y$-component of the pseudospin is flipped by the zigzag edge.
 }
 \label{fig:zigPhase}
\end{figure}

The el-ph matrix element is given as the expectation value 
of the el-ph interaction with respect to the energy eigenstate 
for the unperturbed Hamiltonian,
$H_{\rm K}^0=v_{\rm F} \bsigma \cdot {\bf {\hat p}}$.
The energy eigenstate with wave vector ${\bf k}$
in the conduction energy band is written 
in terms of the plane wave $e^{i{\bf k}\cdot {\bf r}}$ 
and the Bloch function $\Phi^{\rm c}_{\bf k}$ as 
$\Phi^{\rm c}_{\bf k}({\bf r}) =
N e^{i{\bf k}\cdot {\bf r}} \Phi^{\rm c}_{\bf k}$, 
where $N$ is a normalization constant satisfying $N^2 V=1$,
$V$ is the area (volume) of the system, and
\begin{align}
 \Phi^{\rm c}_{\bf k} \equiv \frac{1}{\sqrt{2}}
 \begin{pmatrix}
  1 \cr e^{i\theta({\bf k})}
 \end{pmatrix}.
\label{eq:wf}
\end{align}
Here $\theta({\bf k})$ is the angle 
between the vector ${\bf k}$ and the $k_x$-axis
(see Fig.~\ref{fig:zigPhase}).
The expectation values of 
$\sigma_x$, $\sigma_y$, and $\sigma_z$
with respect to $\Phi^{\rm c}_{\bf k}$
define the pseudospin. 
Since
$\bar{\sigma}_x=\langle \Phi^{\rm c}_{\bf k}| \sigma_x |\Phi^{\rm c}_{\bf k} \rangle
=\cos \theta({\bf k})$,
$\bar{\sigma}_y=\langle \Phi^{\rm c}_{\bf k}| \sigma_y |\Phi^{\rm c}_{\bf k} \rangle
=\sin \theta({\bf k})$,
and 
$\bar{\sigma}_z=\langle \Phi^{\rm c}_{\bf k}| \sigma_z |\Phi^{\rm c}_{\bf k} \rangle
=0$,
the direction of the pseudospin of $\Phi^{\rm c}_{\bf k}$,
\begin{align}
 (\bar{\sigma}_x,\bar{\sigma}_y,\bar{\sigma}_z)=
 (\cos\theta({\bf k}),\sin\theta({\bf k}),0),
\end{align}
is within the $(k_x,k_y)$ plane and 
parallel to the vector ${\bf k}$ (see Fig.~\ref{fig:zigPhase}).
Owing to the presence of the zigzag edge parallel to the $x$-axis,
the wave function near the zigzag edge is a standing wave 
given by a sum of the incident wave $\Phi^{\rm c}_{\bf k}({\bf r})$ 
and the reflected wave 
$\Phi^{\rm c}_{\bf k'}({\bf r})$ with ${\bf k}'\equiv (k_x,-k_y)$ as
\begin{align}
 \Psi^{\rm c}_{\bf k}({\bf r}) = \frac{1}{\sqrt{2}} \left(
 \Phi^{\rm c}_{\bf k}({\bf r}) + \Phi^{\rm c}_{\bf k'}({\bf r})
 \right).
 \label{eq:wfzig}
\end{align}
Strictly speaking, 
it is necessary to add the relative phase between 
$\Phi^{\rm c}_{\bf k}({\bf r})$ and $\Phi^{\rm c}_{\bf k'}({\bf r})$
in order that $\Psi^{\rm c}_{\bf k}({\bf r})$ may satisfy the boundary
condition for the zigzag edge.
However, this phase gives no contribution to the matrix elements of
interest in the present investigation, and therefore we omit it.
Note that the normalization of eq.~(\ref{eq:wfzig})
is adopted for $k_y\ne 0$. 
Some complications arise when $k_y =0$.
For example, when $k_y=0$ and $k_x <0$,
localized wave functions of edge states~\cite{fujita96} 
should be used, which is explained in Appendix\ref{app:a}.

The el-ph matrix element 
from a state in the conduction band to the same state is given as 
$\langle \Psi^{\rm c}_{\bf k}|H^{\rm zig}_{\rm LO/TO} |\Psi^{\rm c}_{\bf k}
 \rangle$.
Using eqs.~(\ref{eq:Hzig}) and (\ref{eq:wfzig}), the pseudospin, and 
$\theta({\bf k}')=-\theta({\bf k})$,
we obtain 
\begin{align}
 & \langle \Psi^{\rm c}_{\bf k}|H^{\rm zig}_{\rm LO} |\Psi^{\rm c}_{\bf k}
 \rangle = 0, 
 \label{eq:zig-LO} \\
 & \langle \Psi^{\rm c}_{\bf k}|H^{\rm zig}_{\rm TO} |\Psi^{\rm c}_{\bf k}
 \rangle =
 v_{\rm F} A^{\rm q}_x \cos \theta({\bf k}).
 \label{eq:zig-TO}
\end{align}
This result shows that the Raman intensity of the LO mode is negligible
compared with that of the TO mode at zigzag edges.
For eq.~(\ref{eq:zig-LO}),
$\langle \Psi^{\rm c}_{\bf k}|\sigma_y |\Psi^{\rm c}_{\bf k} \rangle$
can be rewritten as a sum of two components,
$\langle \Phi^{\rm c}_{\bf k}|\sigma_y |\Phi^{\rm c}_{\bf k} \rangle+
\langle \Phi^{\rm c}_{\bf k'}|\sigma_y |\Phi^{\rm c}_{\bf k'} \rangle$,
since cross terms such as
$\langle \Phi^{\rm c}_{\bf k}|\sigma_y |\Phi^{\rm c}_{\bf k'} \rangle$
vanish.
Because 
$\langle \Phi^{\rm c}_{\bf k}|\sigma_y |\Phi^{\rm c}_{\bf k} \rangle
=\sin\theta({\bf k})$ and 
$\langle \Phi^{\rm c}_{\bf k'}|\sigma_y |\Phi^{\rm c}_{\bf k'} \rangle
=\sin\theta({\bf k'})=-\sin\theta({\bf k})$,
the $y$-component of the pseudospin for the incident wave 
$\Phi^{\rm c}_{\bf k}({\bf r})$ is reflected as shown in Fig.~\ref{fig:zigPhase}.
Thus, we have 
\begin{align}
 \langle \Psi^{\rm c}_{\bf k}|\sigma_y |\Psi^{\rm c}_{\bf k} \rangle =
 0,
 \label{eq:sig_y}
\end{align}
due to the cancellation of the $y$-component of 
the pseudospin between the incident ${\bf k}$-state 
and the reflected ${\bf k'}$-state.
Similarly, in eq.~(\ref{eq:zig-TO}),
$\langle \Psi^{\rm c}_{\bf k}|\sigma_x |\Psi^{\rm c}_{\bf k} \rangle$,
is written as a sum of two components,
$\langle \Phi^{\rm c}_{\bf k}|\sigma_x |\Phi^{\rm c}_{\bf k} \rangle+
\langle \Phi^{\rm c}_{\bf k'}|\sigma_x |\Phi^{\rm c}_{\bf k'} \rangle$.
Because $\langle \Phi^{\rm c}_{\bf k}|\sigma_x |\Phi^{\rm c}_{\bf k} \rangle
=\cos\theta({\bf k})$ and 
$\langle \Phi^{\rm c}_{\bf k'}|\sigma_x |\Phi^{\rm c}_{\bf k'} \rangle
=\cos\theta({\bf k'})=\cos\theta({\bf k})$, we obtain eq.~(\ref{eq:zig-TO}).

\begin{figure}[htbp]
 \begin{center}
  \includegraphics[scale=0.7]{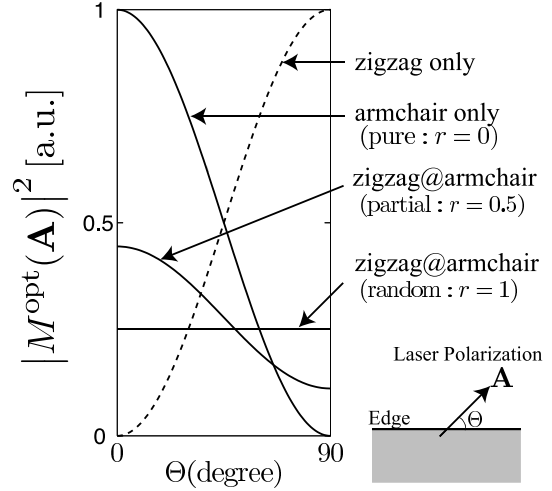}
 \end{center}
 \caption{
 The polarization dependence of the square of 
 the optical transition amplitude ($|M^{\rm opt}({\bf A})|^2$) 
 is plotted as a function of the angle of laser polarization ($\Theta$)
 with respect to the orientation of the edge. 
 For a pure zigzag (armchair) edge, 
 the intensity is maximum when the laser polarization
 is perpendicular (parallel) to the edge.
 ``zigzag@armchair'' denotes the case when 
 zigzag edges are introduced into part of a perfect armchair edge.
 We have used eq.~(\ref{eq:opt-A}) with $r=0$ (armchair only), 
 $r=0.5$ (partial), 
 and $r=1$ (random: a mixture of zigzag and armchair edges).
 }
 \label{fig:Pdepend}
\end{figure}

The Raman intensity depends on the polarization 
of the incident laser light~\cite{grueneis03} 
and on that of the scattered light.
The electron-light interaction 
is given by ${H}^{\rm em}_{\rm K}=-v_{\rm F} e \bsigma \cdot {\bf A}$ in
eq.~(\ref{eq:HK}). 
The optical absorption occurs with amplitude
$M^{\rm opt}({\bf A})=\langle \Psi^{\rm c}_{\bf k}|{H}^{\rm em}_{\rm K}
| \Psi^{\rm v}_{\bf k}\rangle$, where
$\Psi^{\rm v}_{\bf k}({\bf r})$
is the wave function in the valence energy band, 
which is related to $\Psi^{\rm c}_{\bf k}({\bf r})$ via
$\Psi^{\rm v}_{\bf k}({\bf r})= \sigma_z \Psi^{\rm c}_{\bf k}({\bf r})$.
On the other hand, 
the optical emission occurs with amplitude
$\langle \Psi^{\rm v}_{\bf k}|{H}^{\rm em}_{\rm K}| \Psi^{\rm c}_{\bf
k}\rangle$, which is simply the complex conjugate of 
$M^{\rm opt}({\bf A})$.
Thus, the polarization dependences 
of the incident and scattered light are the same. 
Here, let us examine the polarization dependence of the incident light.
The direction of ${\bf A}=(A_x,A_y)$ corresponds to the direction
of the polarization of the electric field.
The polarization of the incident laser light should be perpendicular
to the zigzag edge within a graphene plane, 
i.e., ${\bf A}_\perp=(0,A_y)$,
in order to populate photoexcited electrons effectively.
This argument follows from 
$M^{\rm opt}({\bf A}_\perp) = -i v_{\rm F} e A_y\cos \theta({\bf k})$ 
for ${\bf A}_\perp=(0,A_y)$, while  
$M^{\rm opt}({\bf A}_\parallel) = 0$ for ${\bf A}_\parallel=(A_x,0)$
because 
\begin{align}
 M^{\rm opt}({\bf A}_\perp) &\equiv
 -v_{\rm F} e A_y \langle \Psi^{\rm c}_{\bf k}|\sigma_y |\Psi^{\rm
 v}_{\bf k} \rangle \nn \\
 &= -v_{\rm F} e A_y \langle \Psi^{\rm c}_{\bf k}|\sigma_y \sigma_z
 |\Psi^{\rm c}_{\bf k} \rangle \nn \\
 &= -i v_{\rm F} e A_y \langle \Psi^{\rm c}_{\bf k}|\sigma_x
 |\Psi^{\rm c}_{\bf k} \rangle \nn \\
 &= -i v_{\rm F} e A_y\cos \theta({\bf k}),
 \label{eq:zig-perp}
\end{align}
and 
\begin{align}
 M^{\rm opt}({\bf A}_\parallel) &\equiv
 -v_{\rm F} e A_x \langle \Psi^{\rm c}_{\bf k}|\sigma_x |\Psi^{\rm
 v}_{\bf k} \rangle \nn \\
 &= -v_{\rm F} e A_x \langle \Psi^{\rm c}_{\bf k}|\sigma_x \sigma_z
 |\Psi^{\rm c}_{\bf k} \rangle \nn \\
 &= i v_{\rm F} e A_x \langle \Psi^{\rm c}_{\bf k}|\sigma_y
 |\Psi^{\rm c}_{\bf k} \rangle \nn \\
 &= 0.
 \label{eq:zig-para}
\end{align}
Here, we have used 
$| \Psi^{\rm v}_{\bf k}\rangle = \sigma_z |\Psi^{\rm c}_{\bf k} \rangle$,
$\sigma_y \sigma_z = i\sigma_x$, $\sigma_x \sigma_z = -i\sigma_y$, 
and eq.~(\ref{eq:sig_y}).
It is noteworthy that 
it is mainly the electrons near the $k_x$-axis 
[$\theta({\bf k})\approx 0$ or $\pi$]
that can participate in the Raman process 
taking place near the zigzag edge since both
the el-ph matrix element [eq.~(\ref{eq:zig-TO})]
and the optical transition amplitude [eq.~(\ref{eq:zig-perp})]
are proportional to $\cos\theta({\bf k})$. 
Let us define the angle between the laser polarization 
and the zigzag edge as $\Theta$
(see the inset in Fig.~\ref{fig:Pdepend}), then 
$A_y=|{\bf A}|\sin\Theta$ and 
the Raman intensity is proportional to 
$|M^{\rm opt}({\bf A})|^2 \propto \sin^2 \Theta$.
The $\Theta$-dependence of the square of the optical transition
amplitude is plotted as the dashed curve in Fig.~\ref{fig:Pdepend}.

The Kohn anomaly is relevant to the el-ph matrix element 
for electron-hole pair creation, i.e., 
$\langle \Psi^{\rm c}_{\bf k}|H^{\rm zig}_{\rm LO/TO}| \Psi^{\rm v}_{\bf k}\rangle$. 
Using $\Psi^{\rm v}_{\bf k}({\bf r})= \sigma_z \Psi^{\rm c}_{\bf k}({\bf r})$,
we rewrite the matrix element as
$\langle \Psi^{\rm c}_{\bf k}|H^{\rm zig}_{\rm LO/TO}\sigma_z| \Psi^{\rm
c}_{\bf k}\rangle$. 
From eq.~(\ref{eq:Hzig}), we have
\begin{align}
\begin{split}
 & H^{\rm zig}_{\rm LO} \sigma_z  = iv_{\rm F} A^{\rm q}_y \sigma_x, 
 \\ 
 & H^{\rm zig}_{\rm TO} \sigma_z = -iv_{\rm F} A^{\rm q}_x \sigma_y,
\end{split}
\end{align}
where $\sigma_x \sigma_z = -i\sigma_y$
and $\sigma_y \sigma_z = i\sigma_x$ have been used.
We have thus shown that 
$H^{\rm zig}_{\rm TO} \sigma_z$ is proportional to $\sigma_y$ 
as well as that $H^{\rm zig}_{\rm LO}$
is proportional to $\sigma_y$.
From eq.~(\ref{eq:sig_y}), we see that 
the TO mode is unable to transfer an electron in the valence band 
into the conduction band, that is, 
the TO mode does not decay into an electron-hole pair,
and therefore the Kohn anomaly for the TO mode is 
negligible compared with that for the LO mode.

\section{Armchair Edge}\label{sec:arm}

\begin{figure}[htbp]
 \begin{center}
  \includegraphics[scale=0.4]{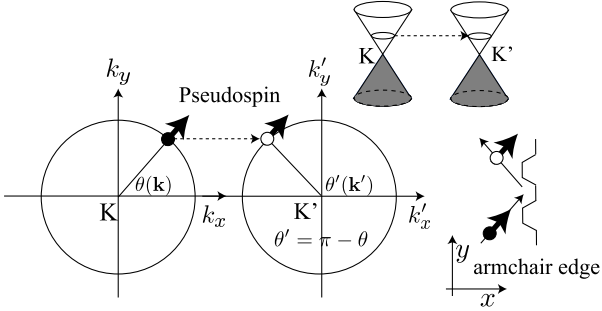}
 \end{center}
 \caption{
 The armchair edge reflects the wave vector 
 ${\bf k}=(k_x,k_y)$ of one valley 
 into ${\bf k'}=(-k_x,k_y)$ of another valley, and
 the two wave functions of the different valleys form a standing wave.
 The pseudospin is unchanged by the armchair edge.
 Note that the pseudospin for states near the K$'$ point 
 is not parallel to the vector ${\bf k'}$, while 
 the pseudospin for states near the K point 
 is parallel to the vector ${\bf k}$.
 }
 \label{fig:armPhase}
\end{figure}

Next, we calculate the matrix element 
relevant to the Raman intensity near the armchair edge.
Suppose that the armchair edge is located along the $y$-axis,
then the armchair edge reflects an electron with ${\bf k}=(k_x,k_y)$ 
near the K point into the state with ${\bf k'}=(k'_x,k'_y)=(-k_x,k_y)$ 
near the K$'$ point, where ${\bf k}$ and ${\bf k'}$
are measured from the K and K$'$ points, respectively.
The negative sign in front of $k_x$ for ${\bf k'}$
is due to the momentum conservation. 
One may consider that 
an intervalley process is unconnected with 
the $\Gamma$ point LO and TO phonons.
Note, however, that we should consider the K and K$'$ points
simultaneously in the case of an armchair edge since 
the reflection of an electron by the armchair edge
is an intervalley scattering process
as shown in Fig.~\ref{fig:armPhase}.

\begin{figure}[htbp]
 \begin{center}
  \includegraphics[scale=0.4]{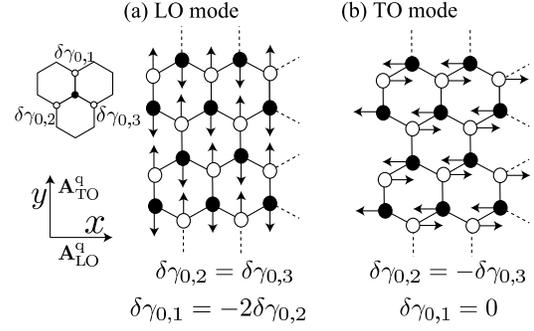}
 \end{center}
 \caption{The displacement vectors for the LO and TO modes 
 are shown in (a) and (b), respectively.
 The displacement vectors of the LO (TO) mode 
 are parallel (perpendicular) to the armchair edge.
 The direction of ${\bf A}^{\rm q}$
 is perpendicular to the direction of atom displacement.
 }
 \label{fig:armLOTO}
\end{figure}

We specify the deformation-induced gauge field 
${\bf A}^{\rm q}$ for the LO and TO modes
near the armchair edge.
The vibrations of carbon atoms 
for the $\Gamma$ point LO and TO modes 
are shown in Fig.~\ref{fig:armLOTO}.
We have $\delta \gamma_{0,1} = -2 \delta \gamma_{0,2}$ and 
$\delta \gamma_{0,2}=\delta \gamma_{0,3}$
for the LO mode, while
$\delta \gamma_{0,1} = 0$ 
and $\delta \gamma_{0,2}=-\delta \gamma_{0,3}$
for the TO mode.
Using eq.~(\ref{eq:gauge}), we see that
${\bf A}^{\rm q}$ for the LO mode
is written as ${\bf A}^{\rm q}_{\rm LO}=(A_x^{\rm q},0)$
with $v_{\rm F}A_x^{\rm q}=-3\delta\gamma_{0,2}$, while 
${\bf A}^{\rm q}$ for the TO mode
is written as ${\bf A}^{\rm q}_{\rm TO}=(0,A_y^{\rm q})$
with $v_{\rm F}A_y^{\rm q}=\sqrt{3}\delta\gamma_{0,2}$.
Thus, from eq.~(\ref{eq:HKK'}),
we see that the el-ph interaction
\begin{align}
 H^{\rm arm}_{\rm LO/TO} =
 v_{\rm F}
 \begin{pmatrix}
  \bsigma \cdot {\bf A}^{\rm q}_{\rm LO/TO} & 0 \cr
  0 & - \bsigma' \cdot {\bf A}^{\rm q}_{\rm LO/TO}
 \end{pmatrix}
\end{align}
is rewritten as
\begin{align}
\begin{split}
 & H^{\rm arm}_{\rm LO} = v_{\rm F} A^{\rm q}_x 
 \begin{pmatrix}
  \sigma_x & 0 \cr 0 & \sigma_x
 \end{pmatrix}, \\ 
 & H^{\rm arm}_{\rm TO} = v_{\rm F} A^{\rm q}_y
 \begin{pmatrix}
  \sigma_y & 0 \cr 0 & - \sigma_y
 \end{pmatrix},
\end{split}
\label{eq:Harm}
\end{align}
for the LO and TO modes, respectively.

The wave function is given by a sum of
the plane wave at the K point and the reflected wave at the K$'$ point as
\begin{align}
 \Psi^{\rm c}_{\bf k}({\bf r}) = \frac{e^{ik_y y}}{\sqrt{2}}
\begin{pmatrix}
 \Phi^{\rm c}_{\bf k} e^{+ik_x x} \cr \Phi^{\rm c}_{\bf k} e^{-ik_x x}
\end{pmatrix}.
 \label{eq:wfarm}
\end{align}
Note that 
the Bloch function is the same ($\Phi^{\rm c}_{\bf k}$)
for both the K and K$'$ points.
In fact, the Bloch function for a state near the K$'$ point 
can be expressed as
\begin{align}
 \frac{1}{\sqrt{2}}
 \begin{pmatrix}
  1 \cr -e^{-i\theta'({\bf k'})}
 \end{pmatrix},
\label{eq:wfK'}
\end{align}
where $\theta'({\bf k'})$ is defined through 
$k'_x+ik'_y=|{\bf k'}|e^{i\theta'({\bf k'})}$.
Since the armchair edge reflects the state with ${\bf k}=(k_x,k_y)$ 
into the state with ${\bf k'}=(-k_x,k_y)$, we have
the relation $\theta'({\bf k'})=\pi-\theta({\bf k})$
(see Fig.~\ref{fig:armPhase}).
By substituting this into eq.~(\ref{eq:wfK'}), 
we see that the Bloch function of eq.~(\ref{eq:wfK'})
becomes $\Phi^{\rm c}_{\bf k}$ of eq.~(\ref{eq:wf}),
which explains eq.~(\ref{eq:wfarm}).
The pseudospin for the eigenstate near the K$'$ point 
is given by 
$\langle \Phi^{\rm c}_{\bf k'}| \sigma_x |\Phi^{\rm c}_{\bf k'} \rangle
=-\cos \theta'({\bf k'})$ and 
$\langle \Phi^{\rm c}_{\bf k'}| \sigma_y |\Phi^{\rm c}_{\bf k'} \rangle
=\sin \theta'({\bf k'})$.
Thus, the pseudospin for the K$'$ point
is not parallel to the vector ${\bf k'}$, 
as shown in Fig.~\ref{fig:armPhase}, 
although the pseudospin for states near the K
point is parallel to the vector {\bf k}.
Using $\theta'({\bf k'})=\pi-\theta({\bf k})$, 
one can see that 
the pseudospin is preserved under the reflection 
at the armchair edge (see Fig.~\ref{fig:armPhase}).

Using eqs.~(\ref{eq:Harm}) and (\ref{eq:wfarm}), 
it is straightforward to check that
\begin{align}
\begin{split}
 & \langle \Psi^{\rm c}_{\bf k}|H^{\rm arm}_{\rm LO} |\Psi^{\rm c}_{\bf k}\rangle 
 = v_{\rm F} A^{\rm q}_x \cos \theta({\bf k}), \\
 & \langle \Psi^{\rm c}_{\bf k}|H^{\rm arm}_{\rm TO} |\Psi^{\rm c}_{\bf k} \rangle 
 = 0.
\end{split}
 \label{eq:arm-LO}
\end{align}
This result shows that 
the Raman intensity of the TO mode is negligible
compared with that of the LO mode.
The absence of the Raman intensity of the TO mode 
results from the interference between two valleys,
namely, the opposite signs in front of $\sigma_y$ 
for the K and K$'$ points of $H^{\rm arm}_{\rm TO}$
in eq.~(\ref{eq:Harm}).

The interaction between 
the light and the electronic states
is given by 
\begin{align}
 {H}^{\rm em}({\bf A}) =
 -v_{\rm F} e
 \begin{pmatrix}
   \bsigma \cdot {\bf A} & 0 \cr
  0 & \bsigma' \cdot {\bf A}
 \end{pmatrix},
\end{align}
from eq.~(\ref{eq:HKK'}).
The optical absorption amplitude is given by 
$M^{\rm opt}({\bf A})=\langle \Psi^{\rm c}_{\bf k}|{H}^{\rm em}({\bf A})
| \Psi^{\rm v}_{\bf k}\rangle$,
where 
\begin{align}
 \Psi^{\rm v}_{\bf k}({\bf r}) = 
 \begin{pmatrix}
  \sigma_z & 0 \cr 0 & \sigma_z
 \end{pmatrix}
\Psi^{\rm c}_{\bf k}({\bf r}).
\end{align}
If the polarization of the incident laser light is perpendicular
to the armchair edge ${\bf A}_\perp=(A_x,0)$, then 
$\langle \Psi^{\rm c}_{\bf k}|{H}^{\rm em}({\bf A}_\perp)|\Psi^{\rm v}_{\bf
k}\rangle$ vanishes owing to the cancellation 
between the K and K$'$ points.
The polarization of the incident laser should be parallel 
to the armchair edge, i.e., ${\bf A}_\parallel=(0,A_y)$,
in order to populate photoexcited electrons effectively
because 
$\langle \Psi^{\rm c}_{\bf k}|{H}^{\rm em}({\bf A}_\parallel)|\Psi^{\rm v}_{\bf
k}\rangle=-iv_{\rm F} eA_y \cos\theta({\bf k})$.
Note that it is mainly the electrons near the $k_x$-axis 
[$\theta({\bf k})\approx 0$ or $\pi$]
that can participate the Raman process 
taking place near the armchair edge since both
the el-ph matrix element [eq.~(\ref{eq:arm-LO})]
and the optical transition amplitude 
are proportional to $\cos\theta({\bf k})$. 
By defining the angle between the laser polarization 
and the armchair edge by $\Theta$
(see the inset in Fig.~\ref{fig:Pdepend}), we have
$A_y=|{\bf A}|\cos\Theta$, and we see that
the Raman intensity is proportional to 
$|M^{\rm opt}({\bf A})|^2 \propto \cos^2 \Theta$.
The polarization dependence of the Raman intensity 
for the armchair edge is opposite that for the zigzag edge,
as shown in Fig.~\ref{fig:Pdepend}, from which 
the orientation of the edge may be determined experimentally.

The el-ph matrix element for the Kohn anomaly is given by 
$\langle \Psi^{\rm c}_{\bf k}|H^{\rm arm}_{\rm LO/TO}\sigma_z| \Psi^{\rm c}_{\bf k}\rangle$. 
From eq.~(\ref{eq:Harm}), we have
\begin{align}
\begin{split}
 & H^{\rm arm}_{\rm LO} \sigma_z = -i v_{\rm F} A^{\rm q}_x 
 \begin{pmatrix}
  \sigma_y & 0 \cr 0 & \sigma_y
 \end{pmatrix}, \\ 
 & H^{\rm arm}_{\rm TO} \sigma_z= i v_{\rm F} A^{\rm q}_y
 \begin{pmatrix}
  \sigma_x & 0 \cr 0 & - \sigma_x
 \end{pmatrix}.
\end{split}
 \label{eq:arm-LO-KA}
\end{align}
It has thus been shown that 
the TO mode does not undergo a Kohn anomaly 
because the matrix element vanishes 
owing to the sign difference between the K and K$'$ points 
with respect to $\sigma_x$.

\section{Energy Difference Between LO and TO Modes}\label{sec:selfene}

In this section 
we calculate the energy difference between the LO and TO modes.
The renormalized phonon energy is written 
as a sum of the unrenormalized energy $\hbar \omega$
and the self-energy.
Since the TO mode does not undergo a Kohn anomaly, 
the self-energy of the TO mode vanishes. 
Thus, the energy difference between the LO and TO modes
is the self-energy of the LO mode, which is given by
time-dependent second-order perturbation theory as
\begin{align}
 \Pi(\omega,E_{\rm F}) = &
 2 \sum_{\bf k} \left(
 \frac{|\langle \Psi^{\rm c}_{\bf k}|H^{\rm arm}_{\rm LO} | \Psi^{\rm
 v}_{\bf k} \rangle|^2}{\hbar \omega -E^{\rm eh}_{\bf k}+i\delta}
 - \frac{|\langle \Psi^{\rm c}_{\bf k}|H^{\rm arm}_{\rm LO} | \Psi^{\rm
 v}_{\bf k} \rangle|^2}{\hbar \omega +E_{\bf k}^{\rm eh}+i\delta}
 \right) \nn \\
 & \times \left(f_{\rm h}-f_{\rm e}\right),
 \label{eq:PI}
\end{align}
where the factor of 2 originates from the spin degeneracy,
$f_{\rm e,h}=(1+\exp((E^{\rm e,h}-E_{\rm F})/k_{\rm B}T)^{-1}$
is the Fermi distribution function,
$E_{\rm F}$ is the Fermi energy, 
$\delta$ is a positive infinitesimal, 
$E^{\rm e}$ ($E^{\rm h}$) is the energy of an electron (a hole),
and $E^{\rm eh}_{\bf k}\equiv E_{\bf k}^{\rm e}-E_{\bf k}^{\rm h}=2\hbar
v_{\rm F}|{\bf k}|$ ($\ge 0$)
is the energy of an electron-hole pair. 
Note that the summation index $\sum_{\bf k}$ in eq.~(\ref{eq:PI}) 
is not restricted to only interband ($E^{\rm eh} \ne 0$) processes
but also includes intraband ($E^{\rm eh}= 0$) processes.
Thus, the self-energy can be decomposed into two parts, 
$\Pi(\omega,E_{\rm F})=\Pi^{\rm inter}(\omega,E_{\rm F})
+\Pi^{\rm intra}(\omega,E_{\rm F})$, where
$\Pi^{\rm inter}(\omega,E_{\rm F})$ includes only 
interband electron-hole pair creation processes 
satisfying $E^{\rm eh}\ne 0$.


In the adiabatic limit, i.e., 
when $\omega = 0$ and $\delta=0$ in eq.~(\ref{eq:PI}),
by substituting eq.~(\ref{eq:arm-LO}) into eq.~(\ref{eq:PI}),
it is straightforward to show that, at $T=0$, 
\begin{align}
\begin{split}
 & \Pi^{\rm intra}(0,E_{\rm F})
 = -\frac{V}{\pi} \left( \frac{A_x^{\rm q}}{\hbar} \right)^2|E_{\rm F}|, \\
 & \Pi^{\rm inter}(0,E_{\rm F})=
 -\frac{V}{\pi} \left( \frac{A_x^{\rm q}}{\hbar} \right)^2 
 \left( E_c - |E_{\rm F}| \right),
\end{split}
\label{eq:ad-pi}
\end{align}
where $E_c$ is a cutoff energy.
Note that $\Pi^{\rm intra}(0,E_{\rm F})$ does not vanish 
because $(f_{\rm h}-f_{\rm e})/E^{\rm eh}_{\bf k} \ne 0$ in the limit of
$E^{\rm eh}_{\bf k}\to 0$, while
in the nonadiabatic case, 
$\Pi^{\rm intra}(\omega,E_{\rm F})$ 
vanishes since $(f_{\rm h}-f_{\rm e})/\hbar \omega=0$ in this limit.
It is only the interband process that contributes to the self-energy 
in the nonadiabatic case.
Lazzeri and Mauri~\cite{lazzeri06prl} 
pointed out that $\Pi(0,E_{\rm F})$ 
does not depend on $E_{\rm F}$
in the adiabatic limit owing to the cancellation
between $\Pi^{\rm intra}(0,E_{\rm F})$ and $\Pi^{\rm inter}(0,E_{\rm
F})$.
This shows that the adiabatic approximation is not appropriate for
discussing the $E_{\rm F}$ dependence of the self-energy.
In the nonadiabatic case, at $T=0$, 
it is a straightforward calculation to obtain
(see Appendix\ref{app:b} for derivation)
\begin{align}
 & {\rm Re} \left[ \Pi(\omega,E_{\rm F}) \right] = \nn \\
 &- \frac{V}{\pi} \left(
 \frac{A_x^{\rm q}}{\hbar} \right)^2 
 \left[ E_c - |E_{\rm F}|- \frac{\hbar \omega}{4}
 \ln \left|  \frac{|E_{\rm F}| - \frac{\hbar \omega}{2}}{|E_{\rm F}|+
 \frac{\hbar \omega}{2} }\right| \right].
 \label{eq:repi}
\end{align}
The Fermi energy dependence is given by the last two
terms.~\cite{lazzeri06prl,ando06-ka} 
The first term is linear with respect to $E_{\rm F}$
and the second term produces a singularity at 
$E_{\rm F}=\pm\hbar \omega/2$. 
These terms express the nonadiabatic effects.~\cite{pisana07}
Recently, Saitta {\it et al.}~\cite{saitta08} have pointed out 
that large nonadiabatic effects are found to be more 
ubiquitous in layered metals such as CaC$_6$ and MgB$_2$.

For the case of $E_{\rm F}=0$, eq.~(\ref{eq:repi}) becomes
\begin{align}
 {\rm Re}\left[\Pi(\omega,0)  \right] =
 - \frac{V}{\pi} \left( \frac{A_x^{\rm q}}{\hbar} \right)^2 E_c.
 \label{eq:pi_inter}
\end{align}
The self-energy depends on the cutoff energy $E_c$.
The value of $E_c$ cannot be determined 
within the effective-mass model.
We assume that 
$E_c$ is of the order of half of the $\pi$ bandwidth (10 eV);
see \S\ref{sec:discussion} for a detailed discussion of the value of
$E_c$. 
Using the harmonic approximation for the displacement 
of the carbon atoms,~\cite{sasaki09} we obtain 
$\sqrt{N_u}|A_x^{\rm q}/\hbar| \approx 2\times 10^{-2}$\AA$^{-1}$
(see Appendix\ref{app:b}),
where $N_u$ denotes the number of hexagonal unit cells. 
Since $V$ can be written as $N_u S$ where $S$ is the area of a hexagonal
unit cell, we obtain ${\rm Re}\left[\Pi(\omega,0)  \right] \approx -6$ meV.
Thus, the difference in the Raman shift between the (Raman active)
TO mode near the zigzag edge and the (Raman active) LO mode near the
armchair edge is approximately 50 cm$^{-1}$.
In a realistic system, the actual magnitude of the self-energy 
may be much smaller than this value.
For example, a typical edge is a mixture of zigzag and armchair edges,
for which the energy difference between the LO and TO modes is lower.

Here, let us introduce zigzag edges 
into part of a perfect armchair edge at $x=0$ 
and examine the effect of the randomness of the edge shape
on the Raman intensity and phonon self-energies.
Then the standing wave near the rough edge is approximated by 
\begin{align}
 \Psi^{\rm c}_{\bf k}({\bf r}) = 
 e^{i{\bf k}\cdot {\bf r}}
 \begin{pmatrix}
  \Phi^{\rm c}_{\bf k} \cr 0
 \end{pmatrix}
 + a e^{i{\bf k'}\cdot {\bf r}}
\begin{pmatrix}
 0 \cr \Phi^{\rm c}_{\bf k} 
\end{pmatrix}
 + z e^{i{\bf k'}\cdot {\bf r}}
 \begin{pmatrix}
  \Phi^{\rm c}_{\bf k'} \cr 0
 \end{pmatrix},
 \label{eq:random}
\end{align}
where ${\bf k'}=(-k_x,k_y)$ and $|a|^2 + |z|^2 = 1$.
The wave function
$\Psi^{\rm c}_{\bf k}({\bf r})$ reproduces eq.~(\ref{eq:wfarm})
for the case when $(a,z)=(1,0)$.
Note that $|z|^2/|a|^2$ ($\equiv r \le 1$) can be considered
phenomenologically as 
the ratio of the number of zigzag edges to that of armchair edges in the
rough edge, and $r=1$ [$(|a|,|z|)=(1/\sqrt{2},1/\sqrt{2})$] 
represents the case that 
armchair and zigzag edges are equally distributed along the $y$-axis.
It is a straightforward calculation to obtain 
\begin{align}
\begin{split}
 & \langle \Psi^{\rm c}_{\bf k}|H^{\rm arm}_{\rm LO}|
 \Psi_{\bf k}^{\rm c} \rangle=v_{\rm F} A_x^{\rm q} \cos\theta({\bf k})
 \frac{1+|a|^2-|z|^2}{1+|a|^2+|z|^2}, \\
 & \langle \Psi_{\bf k}^{\rm c}|H^{\rm arm}_{\rm TO}|\Psi_{\bf k}^{\rm c} \rangle= 
 v_{\rm F} A_y^{\rm q} \sin\theta({\bf k})
 \frac{1-|a|^2+|z|^2}{1+|a|^2+|z|^2}.
\end{split} 
\end{align}
These matrix elements show that 
the self-energy for the LO mode becomes
${\rm Re}\left[\Pi(\omega,0)  \right]/4$ for the case of $r=1$.
On the other hand, 
the self-energy of the TO mode, 
which is zero for the case of $r=0$,
becomes ${\rm Re}\left[\Pi(\omega,0)  \right]/4$ 
for the case of $r=1$.
The differences in the Kohn anomalies
for the LO and TO modes disappear for the case of $r=1$.
Moreover, the Raman intensity of the TO mode increases,
while the Raman intensity of the LO mode decreases. 
As a result, the G band exhibits a single peak. 
The intensity of the G band is given as the sum of the LO and TO modes.
Since the intensity of each mode 
is four times smaller than that of the LO mode 
near the pure armchair edge, the total intensity of the G band 
should be two times smaller than the Raman intensity
near the pure armchair edge.
Note that
for a general value of $(a,z)$, 
the energy difference between the LO and TO modes
is given by
$(|a|^2-|z|^2) {\rm Re}\left[\Pi(\omega,0)  \right]$.
It is also a straightforward calculation to obtain 
the polarization dependence of the optical transition amplitude,
\begin{align}
 |M^{\rm opt}({\bf A})|^2 \propto
 \frac{\cos^2\Theta}{(1+r)^2} + \frac{r^2\sin^2\Theta}{(1+r)^2}.
 \label{eq:opt-A}
\end{align}
This dependence is plotted for two cases, $r=1$ and $r=0.5$,
in Fig.~\ref{fig:Pdepend}.

\section{Bulk and Edge}\label{sec:bulk}

\begin{figure}[htbp]
 \begin{center}
  \includegraphics[scale=0.5]{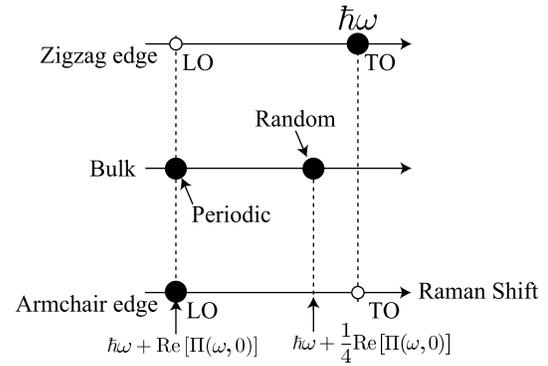}
 \end{center}
 \caption{The horizontal lines indicate the Raman shift for the case of
 $E_{\rm F}=0$.
 (top and bottom lines) The Raman peak taken near the zigzag (armchair)
 edge appears only for the TO (LO) mode indicated by the solid circle.
 The peak for the TO mode does not accompany the broadening 
 because the TO mode decouples from the electron-hole pairs.
 (middle line) The Raman peak appears at $\hbar \omega +{\rm Re}\left[\Pi(\omega,0)  \right]$
 in the case of an infinite periodic graphene system without an edge
 (``Periodic''). If the effect of the electron reflection at the edge 
 survives in the interior of a graphene sample (``Random''), 
 the Raman peak is expected to appear at $\hbar \omega +{\rm Re}\left[\Pi(\omega,0)  \right]/4$.
 }
 \label{fig:spectrum}
\end{figure}

In the case of an infinite periodic graphene system without an edge,
the self-energies of the LO and TO modes are the same and given 
by ${\rm Re}\left[\Pi(\omega,E_{\rm F})  \right]$ in eq.~(\ref{eq:repi}).
Moreover, no asymmetry between the LO and TO modes in the Raman
intensity is expected.
The reason why 
the LO and TO modes do not exhibit any difference in Raman spectra
is that 
graphene is a homo-polar crystal with two atoms per unit cell, and hence 
there is no polar mode, similar to the case of Si.
Thus, the LO and TO modes are degenerate and contribute equally 
to the single peak of the G band 
(see ``Periodic'' in Fig.~\ref{fig:spectrum}). 
Note that a slight change in the spring force constant 
due to a uniaxial strain applied to a graphene sample
can resolve the degeneracy between the LO and TO modes. 
In this case, the unrenormalized energy $\hbar \omega$ for the LO mode is
not identical to that for the TO mode. 
However, even for this case, we can expect that the self-energies and Raman
intensities for the LO and TO modes are similar to each other.
Thus, we can see two peaks for the LO and TO modes 
with similar intensity, as was observed by Mohiuddin {\it et al.}~\cite{mohiuddin09}

Since an actual sample is always surrounded by an edge,
it is interesting to consider whether or not 
the interior of a graphene sample can be considered 
as an infinite periodic graphene system without the edge.
If the wave function in the interior region is given by 
a superposition of the incident and reflected states, 
then it is reasonable to assume that the wave function is 
approximated by eq.~(\ref{eq:random}) with 
$(|a|,|z|)=(1/\sqrt{2},1/\sqrt{2})$,
since it is probable that the edge is a random mixture of 
zigzag and armchair edges. 
The peak positions of the LO and TO modes in the Raman shift 
are indicated by ``Random'' in Fig.~\ref{fig:spectrum}.
We speculate that the peak position for an actual sample 
appears between the peaks labeled ``Periodic'' and ``Random''.
An estimation of the effective distance from the edge 
at which the effect of interference on the pseudospin
discussed so far can survive will be a subject of further
investigation.

\section{Discussion and Conclusions}\label{sec:discussion}

Here,
we discuss the relationship between our result and experimental results.
Can\ifmmode \mbox{\c{c}}\else \c{c}\fi{}ado {\it et al}. 
observed that the Raman intensity of the G band for a nanoribbon
has a strong dependence on the incident light polarization.~\cite{canifmmode04}
They showed that the Raman intensity is maximum when 
the polarization is parallel to the edge of a nanoribbon. 
Their result is consistent with our result for the armchair edge, 
but not consistent with our result for the zigzag edge. 
We speculate that the sample used in their experiment
is similar to a nanoribbon with an armchair edge. 
This speculation is reasonable
because armchair edges are more frequently observed in experiments 
than zigzag edges.~\cite{kobayashi05}
Casiraghi {\it et al}. performed Raman spectroscopy on graphene edges
and observed a small redshift of the G peak near the edge accompanied
by a decrease in the linewidth of the G peak.~\cite{casiraghi09}
This behavior of the G peak is consistent with that of the zigzag edge
since it is only the TO mode without broadening (which is related to the
imaginary part of the self-energy) that can be Raman active.

The cutoff energy $E_c$ 
appearing in eq.~(\ref{eq:pi_inter})
may be determined from a tight-binding lattice model.
For periodic graphene, by taking into account the contribution of all
the possible electron-hole intermediate states in the Brillouin zone, 
we can have $E_c \approx 7\gamma_0$ (20 eV),
which is larger than the value adopted in eq.~(\ref{eq:pi_inter}).
The value of $E_c$ for a graphene sample with an edge 
may be different from that for a periodic graphene sample without an edge. 
In fact, for a nanoribbon, 
a tight-binding calculation~\cite{sasaki09} shows that 
the energy difference between the LO and TO modes 
is approximately 30 cm$^{-1}$,
which corresponds to $E_c\approx 6$ eV.
Thus the value of $E_c$ depends on the geometry of the system.
Since we have considered a large graphene sample with an edge,
we assumed that an appropriate value of $E_c$ is 
between 6 and 20 eV, and we chose 10 eV,
which is of the order of half of the $\pi$ bandwidth.
Because $E_c$ is not an experimentally controllable parameter, 
we consider that, in order to verify our results, 
it is essential to observe 
the $E_{\rm F}$ dependence of the G band spectra
near the edge.

In conclusion, 
the el-ph matrix elements 
for the Raman intensity and Kohn anomaly 
near the edge of graphene were derived by adiabatic calculation, 
and then perturbation treatment was applied to 
the nonadiabatic parts of the phonon self-energies. 
The zigzag edge causes intravalley scattering
and the $y$-component of the pseudospin vanishes 
$\langle \sigma_y \rangle = 0$ for the standing wave.
The Raman intensity of the LO mode and the Kohn anomaly of the TO mode
are negligible owing to $\langle \sigma_y \rangle = 0$.
On the other hand, 
the armchair edge causes intervalley scattering
and the pseudospin does not change its direction. 
However, 
owing to the interference between two valleys 
originating from the el-ph interaction,
the Raman intensity and Kohn anomaly are negligible only for the TO
mode.  
The Raman intensity is enhanced 
when the polarization of the incident laser is parallel (perpendicular)
to the armchair (zigzag) edge. 
The difference in the behavior of the pseudospin 
with respect to the zigzag and armchair edges is the origin of the
asymmetry between the LO and TO modes.
Our results are summarized in Table~\ref{tab:1}.

\begin{table}[htbp]
 \caption{\label{tab:1}
 Dependences of the Raman intensities and Kohn anomalies
 on the $\Gamma$ point optical phonon modes.
 The symbols $\bigcirc$ and $\times$ for Raman intensity and the Kohn
 anomaly represent `occurrence' and `absence', respectively. 
 There is asymmetry between the Raman intensity and Kohn anomaly, that is,
 the Kohn anomaly occurs only for the LO mode, while the mode with a
 strong Raman intensity changes according to the edge shape.
 Raman intensity is enhanced when the polarization of the incident laser
 light is parallel (LO) to the armchair edge
 or when it is perpendicular (TO) to the zigzag edge.
 }
  \begin{tabular}{c|cccc}
  {\bf Position} & {\bf Mode} & {\bf Raman} & {\bf Kohn} & {\bf
   Polarization}\\
   \hline 
   {\bf zigzag} & LO & $\times$ & $\bigcirc$ & $\times$ \\
        & TO & $\bigcirc$ & $\times$ & $\bigcirc$  \\ 
   \hline
   {\bf armchair} & LO & $\bigcirc$ & $\bigcirc$ & $\bigcirc$ 
   \\ & TO & $\times$ & $\times$ & $\times$ \\
   \hline
   {\bf bulk} & LO & $\bigcirc$ & $\bigcirc$ & $\bigcirc$ 
   \\ & TO & $\bigcirc$ & $\bigcirc$ & $\bigcirc$ \\
   \end{tabular}
\end{table}

\section*{Acknowledgments}

K. S. would like to thank T. Osada (Institute for Solid State Physics,
University of Tokyo) for a useful comment on the pseudospin for states
near the K$'$ point. 
R. S. acknowledges a MEXT Grant (No.~20241023).
This work was supported by 
a Grant-in-Aid for Specially Promoted Research
(No.~20001006) from MEXT.

\appendix

\section{Correction of Edge States to eq.~(\ref{eq:zig-TO})}\label{app:a}

Here, we exactly calculate the $x$-component of the pseudospin,
$\langle \Psi^{\rm c}_{\bf k}|\sigma_x |\Psi^{\rm c}_{\bf k} \rangle$,
with $k_y=0$.
When $k_y \to 0$, we have
\begin{align}
 \langle \Psi^{\rm c}_{\bf k}|\sigma_x |\Psi^{\rm c}_{\bf k} \rangle
 = \cos\theta({\bf k}).
 \label{app:1}
\end{align}
It should be noted that this expression holds for extended states.
The states with $k_y=0$ 
are divided into two states, extended states and edge states,~\cite{fujita96}
depending on the sign of $k_x$.~\cite{sasaki06jpsj}
For the K point, the edge states satisfy $k_x<0$, 
while the extended states satisfy $k_x>0$.
Since the edge states are pseudospin polarization states, that is, they
are eigenstates of $\sigma_z$, then the matrix element of $\sigma_x$ 
with respect to the edge states vanishes. 
Thus, the exact form is given by 
\begin{align}
 \langle \Psi^{\rm c}_{\bf k}|\sigma_x |\Psi^{\rm c}_{\bf k} \rangle
 = \left\{
 \begin{array}{@{\,}ll}
  \cos \theta({\bf k}) & (k_y \ne 0), \\
  {\rm sign}(k_x)\cos \theta({\bf k}) & (k_y = 0),
 \end{array} \right.
\end{align}
where ${\rm sign}(k_x)=1$ when $k_x>0$ and zero otherwise.
By neglecting this complication, we obtain eq.~(\ref{eq:zig-TO}).

\section{Derivation of eq.~(\ref{eq:repi})}\label{app:b}

In this section, we derive eq.~(\ref{eq:repi}).

First, using eqs.~(\ref{eq:wfarm}) and (\ref{eq:arm-LO-KA}),
we obtain 
\begin{align}
 \langle \Psi^{\rm c}_{\bf k}|H^{\rm arm}_{\rm LO} | \Psi^{\rm v}_{\bf k}
 \rangle=-iv_{\rm F}A_x^{\rm q}\sin \theta({\bf k}). 
 \label{app:f2}
\end{align}
Next, we consider the real part of the self-energy 
by setting $\delta = 0$ in eq.~(\ref{eq:PI}).
At zero temperature, 
we can set $f_{\rm h}-f_{\rm e}=1$ 
for $E^{\rm e}_{\bf k}\ge |E_{\rm F}|$, otherwise $f_{\rm h}-f_{\rm e}=0$.
Then, the self-energy of the LO mode is written as
\begin{align}
 {\rm Re} \left[ \Pi(\omega,E_{\rm F}) \right] =  2 (v_{\rm F}A_x^{\rm
 q})^2 \sum'_{\bf k} 
  \left\{
 \frac{\sin^2 \theta({\bf k})}{\hbar \omega-E_{\bf k}^{\rm eh}}
 -\frac{\sin^2 \theta({\bf k})}{\hbar \omega+E_{\bf k}^{\rm eh}}\right\},
 \label{app:f3}
\end{align}
where $\sum'$ indicates that the summation is taken over states 
satisfying $E^{\rm e}_{\bf k}\ge |E_{\rm F}|$.
Since the $y$-axis ($x$-axis) is parallel (perpendicular) to the armchair
edge, we use a periodic boundary condition for $k_y$ and 
an open boundary condition for $k_x$. 
Then we have $k_y=2\pi n_y/L_y$ and $k_x = \pi n_x /L_x$.
The summation over possible electron-hole pairs 
can be rewritten as 
\begin{align}
 \sum_{\bf k} = 2 \left[ \frac{V}{2\pi^2} 
 \int_0^{k_c} dk_x \int_{-k_c}^{k_c} dk_y \right],
 \label{app:f1}
\end{align}
where $V\equiv L_x L_y$, $k_c$ is the cutoff momentum, and 
the factor of 2 originates from the degeneracy with respect to the K and
K$'$ points. 
Substituting eq.~(\ref{app:f1}) into eq.~(\ref{app:f3}),
we have
\begin{align}
 {\rm Re} \left[ \Pi(\omega,E_{\rm F}) \right] &=  
 \frac{2V}{\pi^2} \left( \frac{A_x^{\rm q}}{\hbar}\right)^2 
 \int_{-\pi/2}^{\pi/2} \sin^2 \theta d\theta  \nn \\
 & \times \int_{|E_{\rm F}|}^{E_c} EdE
  \left\{
 \frac{1}{\hbar \omega-2E}
 -\frac{1}{\hbar \omega+2E}\right\},
\end{align}
where we have changed the integration variables from $(k_x,k_y)$
to $(E,k)$ by using $k_x = k\cos \theta$, $k_y=k\sin\theta$, and
$E=\hbar v_{\rm F}k$.
Using $\int_{-\pi/2}^{\pi/2} \sin^2 \theta d\theta=\pi/2$ and 
$\int x/(x+a) dx=x-a\ln|x+a|$,
we obtain eq.~(\ref{eq:repi}) when $E_c \gg \hbar \omega$.

To calculate $A_x^{\rm q}$, we have used 
$A_x^{\rm q} \equiv g_{\rm off} u(\omega)/a_{\rm cc}$
where $a_{\rm cc}=1.42$ \AA.
Here $g_{\rm off}$ is the off-site el-ph matrix element 
and $u(\omega)$ is the amplitude of the phonon mode.
We adopt $g_{\rm off} = 6.4$ eV.~\cite{jiang05prb} 
A similar value is obtained by a first-principles calculation 
with the local density approximation.~\cite{porezag95}
We use a harmonic oscillator model which gives 
$u(\omega)=\sqrt{\hbar/2M_c N_u \omega}$, where
$M_c$ is the mass of a carbon atom.
Using $\hbar \omega=0.2$ eV, we obtain
$\sqrt{N_u}|A_x^{\rm q}/\hbar| \approx 2\times 10^{-2}$\AA$^{-1}$.


\end{document}